\documentclass[fleqn,10pt]{wlscirep}
\pdfoutput=1  

\usepackage{upgreek}

\title{Cooling the Motion of Diamond Nanocrystals in a Magneto-Gravitational Trap in High Vacuum}

\author[1]{Jen-Feng Hsu}
\author[1]{Peng Ji}
\author[1]{Charles W. Lewandowski}
\author[1,*]{Brian D'Urso}
\affil[1]{Department of Physics and Astronomy, University of Pittsburgh, Pittsburgh, PA 15260, USA}

\affil[*]{dursobr@pitt.edu}

\begin{abstract} 
Levitated diamond nanocrystals with nitrogen-vacancy (NV) centres in high vacuum have been proposed as a unique system for experiments in fundamental quantum mechanics, including the generation of large quantum superposition states and tests of quantum gravity. This system promises extreme isolation from its environment while providing quantum control and sensing through the NV centre spin. While optical trapping has been the most explored method of levitation, recent results indicate that excessive optical heating of the nanodiamonds under vacuum may make the method impractical with currently available materials. Here, we study an alternative magneto-gravitational trap for diamagnetic particles, such as diamond nanocrystals, with stable levitation from atmospheric pressure to high vacuum. Magnetic field gradients from permanent magnets confine the particle in two dimensions, while confinement in the third dimension is gravitational. We demonstrate that feedback cooling of the centre-of-mass motion of a trapped nanodiamond cluster results in cooling of one degree of freedom to less than 1~K.
\end{abstract}

\begin{document}

\flushbottom
\maketitle
%
%
\thispagestyle{empty}


\section*{Introduction}

Extreme isolation from the surrounding environment makes optically trapped particles in vacuum an attractive system in which to study fundamental quantum mechanics~\cite{chang2010cavity,li2011millikelvin,kiesel2013cavity}. If the trapped particle is a diamond nanocrystal with nitrogen-vacancy (NV) defect centres, the spin of the NV centres can be manipulated~\cite{doherty2013nitrogen} to provide a quantum handle into the system. The spin state can be used to alter the magnetic moment of the trapped particle and spectroscopy on spin transitions can measure the local magnetic field~\cite{lai2010optical,horowitz2012electron}. A diamond nanocrystal with NV centres levitated in an optical trap has been proposed as a platform for experiments including generating macroscopic superposition (Schr\"{o}dinger's cat) states~\cite{yin2013large,scala2013matter} and tests of quantum gravity~\cite{kafri2014classical,albrecht2014testing}. 

Trapping in vacuum is critical for decreasing the natural damping of the mechanical motion and opening the possibility of cooling the mechanical motion to near the quantum ground state. However, there is a gap in experiments with optically trapped diamond nanocrystals: trapping at high vacuum has not been demonstrated, although multiple experiments show diamond nanocrystals trapped at higher pressures~\cite{neukirch2013observation,li2016observation}. Recent evidence shows that some nanodiamonds burn or graphitize in an optical trap whenever the pressure is less than $10$~mbar~\cite{rahman2016burning}, providing a likely explanation for the challenge. While it may be possible to reduce the absorption of the trapping laser by using higher purity nanodiamonds, a more robust solution would be to levitate nanodiamonds with an alternative trap which does not require high intensity oscillating electromagnetic fields.

In this report, we experimentally demonstrate the use of a combination of static magnetic field gradients and gravity for stable trapping of a diamond nanocrystal cluster in high vacuum. The weak diamagnetism of diamond provides the force needed to levitate the nanodiamonds~\cite{simon2000diamagnetic}. We analyse the potential of nanodiamonds in the trap, which results in independent harmonic motion in three directions. Unlike previous experimental studies of systems with diamagnetic levitation~\cite{lyuksyutov2004chip,pigot2008diamagnetic,hill2012shape,gunawan2015parallel}, we demonstrate trapping in vacuum, characterise the harmonic centre-of-mass motion of trapped particles, and cool the motion by up to a factor of $490$ using position detection and feedback. Unlike other proposed quantum magnetomechanical systems~\cite{romero2012quantum,cirio2012quantum}, we do not require superconducting particles for levitation. The diamagnetism of many common materials, such as diamond, is adequate for levitation with a strong magnetic field gradient. Since the trap utilises only static fields, there is no intrinsic heating of the particle, making it an attractive alternative to optical traps. The low stiffness and large size scale of the trap (relative to optical traps) result in low oscillation frequencies of $10$ to $150$~Hz, but also make detection simple. For example, the motion of a particle when equilibrated at ambient temperature in the trap can be easily observed and recorded with a high-speed camera.

\section*{Trap Description}

\begin{figure}[ht]
\centerline{\includegraphics[width=1\columnwidth]{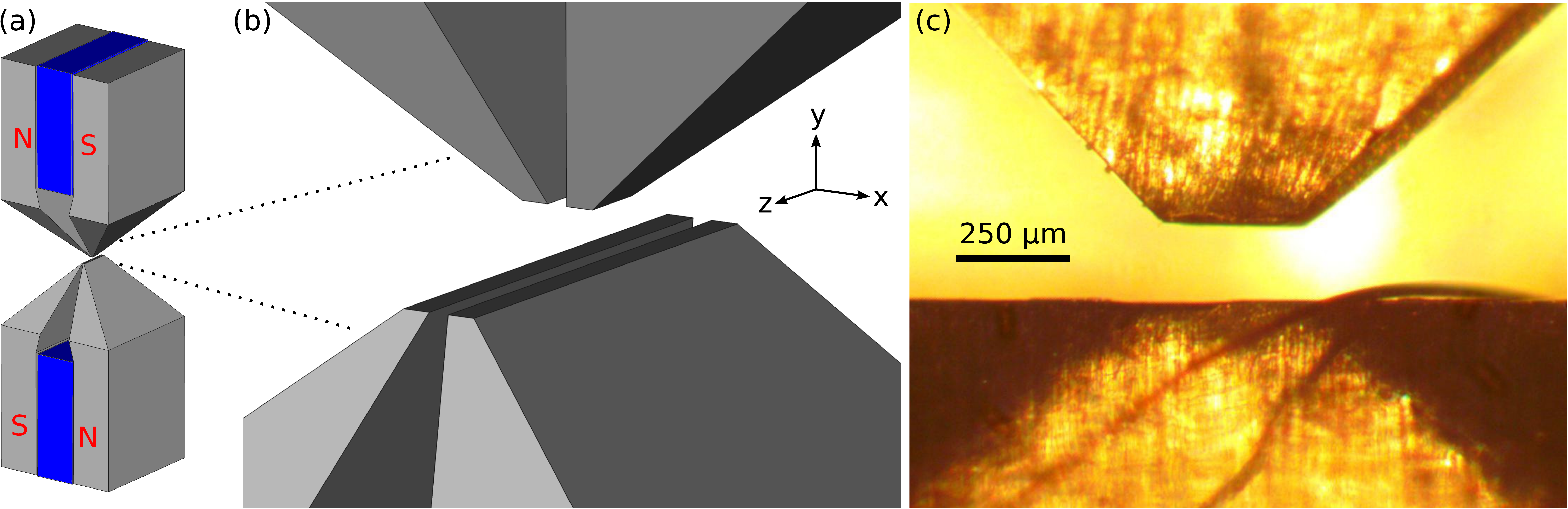}}
\caption{Images of the magneto-gravitational trap. (a) Illustration of the trap with FeCo (Hiperco 50A Alloy, Carpenter Technology) pole pieces (grey) and SmCo magnets (blue). (b) Closer view of the pole pieces near the trapping region and the chosen coordinate system. (c) Photograph of the trapping region from the transverse ($x$) direction, also showing the wire running through the trap for cooling the mechanical motion of a trapped particle.}
\label{fig:trap_illustration}
\end{figure}

The magneto-gravitational trap, which utilises two permanent SmCo magnets and four FeCo alloy pole pieces to concentrate and direct the field, is shown in Fig.~\ref{fig:trap_illustration}. The shape of the resulting magnetic field and the trapping mechanism in each direction can be understood from a multipole expansion of the field. We consider only the three lowest non-zero multipole moments, since we can determine three expansion coefficients from the three measurable centre-of-mass oscillation frequencies of a trapped particle (for full derivation, see Supplementary Information). 

\begin{figure}[ht]
\centerline{\includegraphics[width=1\columnwidth]{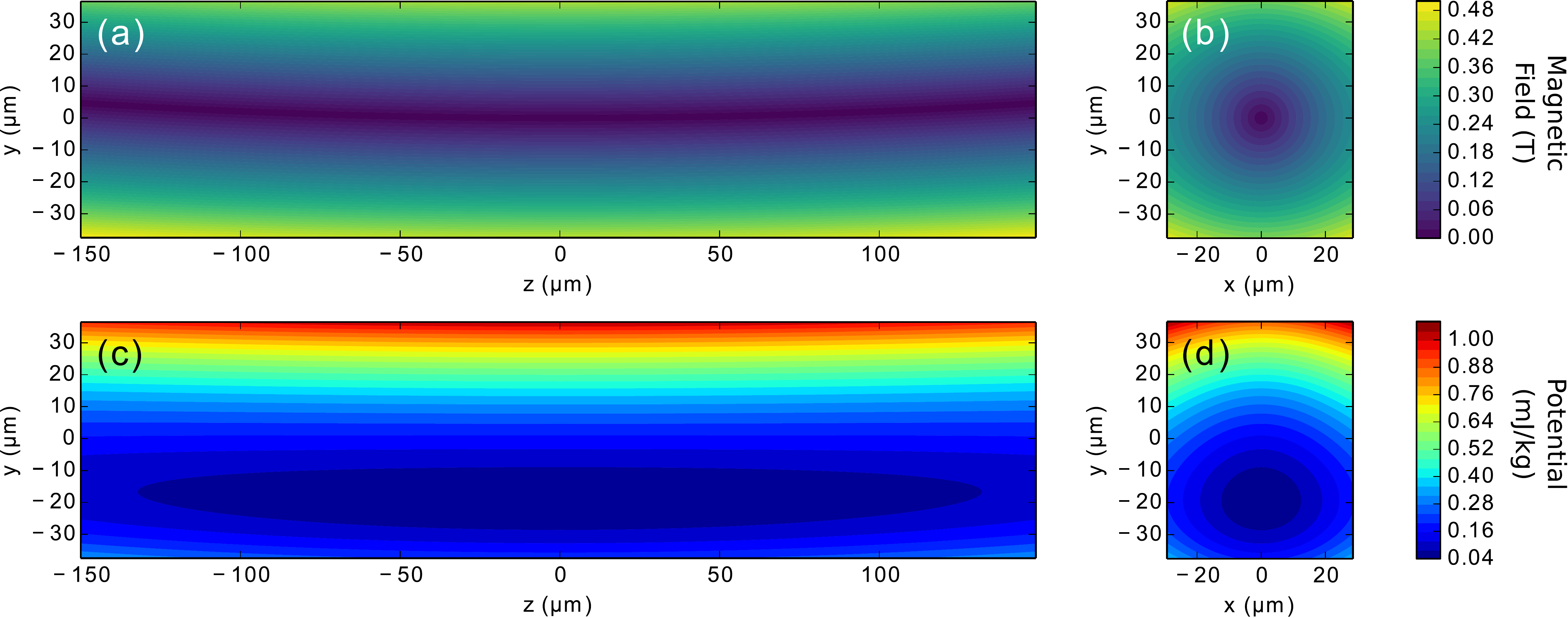}}
\caption{Plots of (a) and (b) the magnitude of the magnetic field and (c) and (d) the total potential energy per unit mass of a diamond nanocrystal in the trap, with contributions from the magnetic field and gravity. (a) and (c) show cross-sections through the $y$-$z$ plane, while (b) and (c) show cross-sections through the $x$-$y$ plane.}
\label{fig:multipole_expansion}
\end{figure}

The trap is based around a linear magnetic quadrupole field, which can be described by a magnetic scalar potential 
\begin{equation}
\Phi_{2,-2}\left(x, y, z\right) = 
\left(\frac{a_2 y_0}{2\mu_0}\right)\left(\frac{r}{y_0}\right)^2
\left(\frac{1}{2} \sqrt{\frac{15}{\pi}} \frac{xy}{r^2}\right),
\end{equation}
where $\mu_0$ is the permeability of free space, $r=\sqrt{x^2+y^2+z^2}$ is the distance from the geometrical centre of the trap, $y_0=75$~$\upmu$m is the distance from the centre of the trap to the top or bottom pole pieces, and $a_2$ is a coefficient that sets the strength of the quadrupole. This magnetic scalar potential represents a magnetic field which is zero on the $z$ axis, increases in the $y$ (vertical) and $x$ (transverse) directions, and is invariant under translations in the $z$ (axial) direction. Since experimentally the pole pieces must be finite in length and even an ideal linear quadrupole field does not constrain the particle in all directions, we cut the pole pieces to finite length in the axial direction and break vertical symmetry by making the top pole pieces shorter than the bottom pole pieces. With this change, the line of zero magnetic field curves vertically upward as $z$ increases or decreases from zero (see Fig.~\ref{fig:multipole_expansion}(a)). This modification to the quadrupole field can be described by the addition of a hexapole magnetic scalar potential 
\begin{equation}
\Phi_{31}\left(x, y, z \right) = 
\left(\frac{a_3 y_0}{3\mu_0}\right)\left(\frac{r}{y_0}\right)^3
\left(\frac{1}{4} \sqrt{\frac{21}{2\pi}} \frac{x(4z^2-x^2-y^2)}{r^3}\right),
\end{equation}
where $a_3$ is a coefficient that sets the strength of the hexapole. A final complication of the experimental trap is that the pole pieces are not distributed with four-fold symmetry; they are polished flat horizontally and the top and bottom pairs are each brought closer together to increase the available numerical aperture for light collection in the transverse direction. These modifications also strengthens the magnetic field at the top and bottom of the trap, making it better able to oppose the pull of gravity in the negative $y$ direction (see Fig.~\ref{fig:multipole_expansion}(b)). This modification can be described by the addition of a linear octopole magnetic scalar potential
\begin{equation}
\Phi_{4,-4}\left(x, y, z \right) = 
\left(\frac{a_4 y_0}{4\mu_0}\right)\left(\frac{r}{y_0}\right)^4
\left(\frac{3}{4} \sqrt{\frac{35}{\pi}} \frac{xy(x^2-y^2)}{r^4}\right),
\end{equation}
where $a_4$ is a coefficient that sets the strength of the octopole. While higher multipoles could be included in the field description, these three capture the most critical features of the trap. The total magnetic field is
\begin{equation}
\vec{B}\left(x, y, z \right) \approx -\mu_0\vec{\nabla}\left(
\Phi_{2,-2}\left(x, y, z \right) 
+ \Phi_{31}\left(x, y, z \right) 
+ \Phi_{4,-4}\left(x, y, z \right)
\right).
\end{equation}

Since the diamagnetic trapping force is weak, the force of gravity cannot be neglected. Furthermore, gravity prevents the particle from drifting out of the trap described above in the axial direction and shifts the equilibrium position of the particle vertically downward, so it does not tend to rest in the zero-field region. Including gravity, the potential energy of a particle in the trap is
\begin{equation}
U = -\frac{\chi B^2 V}{2\mu_0} + m g y,
\end{equation}
where $\chi$ is the magnetic susceptibility of the particle ($<0$ for diamagnetic materials), $V$ is the volume of the particle, $m$ is the mass of the particle, and $g$ is the acceleration due to gravity, as plotted in Fig.~\ref{fig:multipole_expansion}(c) and (d). For the diamagnetic force to balance gravity on a small particle, we must have~\cite{simon2000diamagnetic}
\begin{equation}
m g = \frac{\chi V}{2\mu_0} \left. \frac{d}{dy}B^2 \right|_{y=y_\mathit{eq}} 
\qquad\mathrm{or}\qquad
\rho g = \frac{\chi}{2\mu_0} \left. \frac{d}{dy}B^2 \right|_{y=y_\mathit{eq}},
\label{eq:levitation}
\end{equation}
where $\rho$ is the mass density of the particle material and $y_\mathit{eq}$ is the equilibrium $y$ position. For small particles, the oscillation frequencies are
\begin{equation}
\omega_\xi = \left.\sqrt{\frac{1}{m}\frac{d^2 U}{d\xi^2}}\right|_{x=0,y=y_\mathit{eq},z=0}
=\left.\sqrt{\frac{-\chi}{2\rho\mu_0} \frac{d^2B^2}{d\xi^2}}\right|_{x=0,y=y_\mathit{eq},z=0},
\label{eq:frequencies}
\end{equation}
where $\xi=x$, $y$, or $z$ for the direction of oscillation. Note that the oscillation frequencies (Eq.~\ref{eq:frequencies}) and the levitation condition (Eq.~\ref{eq:levitation}) depend on the material properties of the particle and the magnetic field geometry, but are independent of the size of the particle. Therefore, different particles of the same materials will have similar behaviour in the trap.

The coefficients in the multipole expansion of the magnetic field can be matched to the experimentally measured trap frequencies by setting $x = z = 0$ and solving for  $y_\mathit{eq}$, $a_2$, $a_3$, and $a_4$ while requiring that the net force on the particle is zero and that the predicted trap frequencies match the experimental values. Results for the trapped diamond nanocrystal cluster presented here are $\omega_x=2\pi\times 104$~Hz,  $\omega_y=2\pi\times 130$~Hz, $\omega_z=2\pi\times 9.6$~Hz, $y_\mathit{eq}=-19$~$\upmu$m, $a_2=-1.3$~T, $a_3=0.018$~T, and $a_4=0.72$~T. The resulting magnetic field and total potential energy in the trap are illustrated in Fig.~\ref{fig:multipole_expansion}. Due to gravity, the magnitude of the magnetic field at the equilibrium position of trapped diamond nanocrystals is not zero, instead it is $B_\mathit{eq}\approx 0.2$~T. Such a large magnetic field could cause problems with electron spin resonance (ESR) measurements on NV centres in the trapped nanodiamond, particularly if it is not aligned to the NV centre axis. While it is beyond the scope of this report, we have found that it is possible to use an electric field to raise the equilibrium position of a particle in the trap, so this challenge is not insurmountable.

\begin{figure}[ht]
\centerline{\includegraphics[width=1\columnwidth]{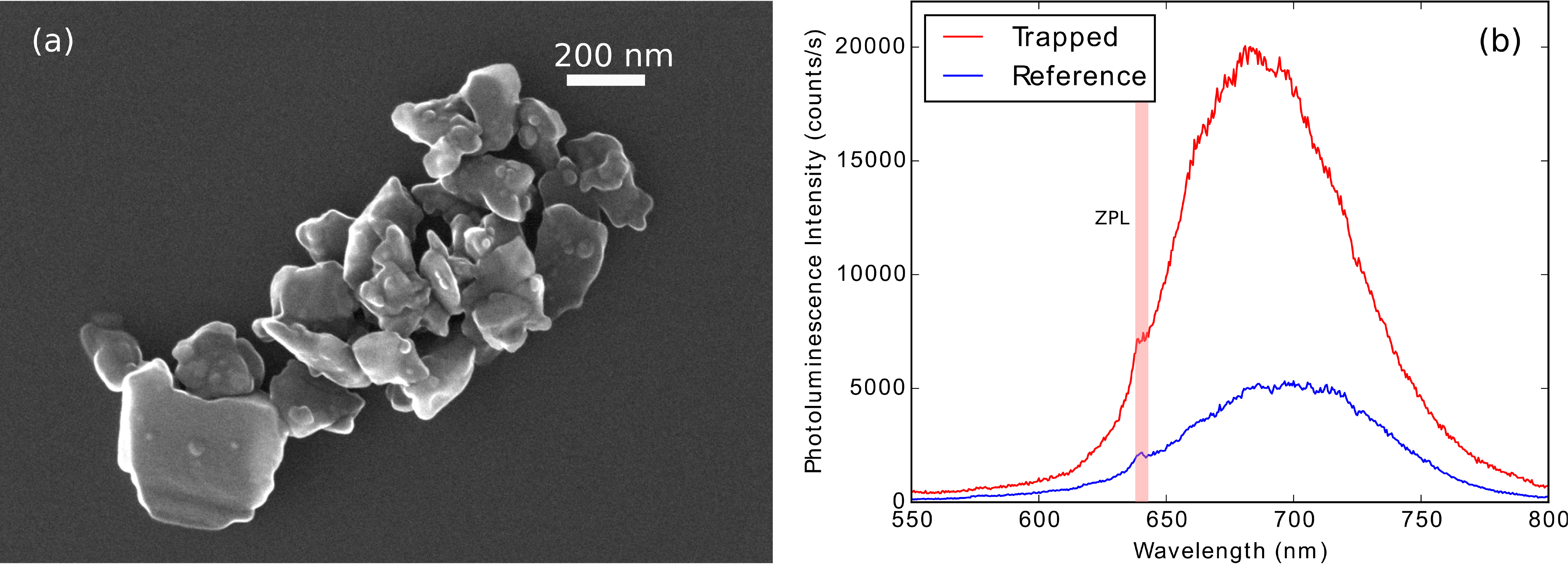}}
\caption{(a) Scanning electron microscope (SEM) image of diamond nanocrystals on a solid substrate. (b) Photoluminescence spectra of a trapped diamond nanocrystal cluster and a reference spectrum of diamond nanocrystals on a silicon substrate taken through the same optical setup, with the location of the ZPL indicated. The integration time for the trapped particle was $0.6$~s per point with peak pump  intensity $\approx 35$~nW/$\upmu$m$^2$, while the reference spectrum used a longer integration time per point ($1.8$~s) and a higher peak pump intensity ($\approx 600$~nW/$\upmu$m$^2$) due to poor coupling of the pump from the axial direction.}
\label{fig:nanodiamonds}
\end{figure}

\section*{Loading and Identifying Diamond Nanocrystals}
Particles are loaded into the magnetic trap using a technique similar to that for loading optical traps~\cite{neukirch2013observation}. First, diamond nanocrystals (Ad\'amas Nanotechnologies ND-1600NV-140nm in isopropyl alcohol, see Fig.~\ref{fig:nanodiamonds}(a)) are suspended in additional isopropyl alcohol and dibutyl sebacate (DBS). Droplets of the suspension are sprayed into the trap region with an ultrasonic horn\cite{perron1967design} at atmospheric pressure, where the isopropyl alcohol rapidly evaporates. DBS evaporates slowly, giving time to pump the surrounding vacuum chamber to the desired pressure using mechanical and turbomolecular pumps. Once high vacuum ($< 10^{-3}$~mbar) is reached, the mechanical motion of the droplet may be cooled as described in the next section. To minimise electrical charge on the drop, a source of ionising radiation (Americium-241) is placed in the chamber near the trap to supply ions which neutralise the charge on the particle after loading and during pumping. Once trapped and under vacuum, particles can be held in the trap indefinitely. If multiple droplets are trapped initially, they typically coalesce once electrically neutralised and ultimately dry to a single cluster of diamond nanocrystals. 

To check the identity of the particles, the photoluminescence spectrum of the trapped particles is measured while pumping the NV centres with $520$~nm laser light from the axial direction (see Fig.~\ref{fig:optical_setup}). Since the scattering force from the pump beam (with peak intensity $\approx 35$~nW/$\upmu$m$^2$) can easily drive the particle out of the trap, we pulse the pump beam and cool the mechanical motion of the particle back to near equilibrium after each excitation pulse ($5$~ms pump pulses, $2$~Hz repetition rate). A recorded spectrum is shown in Fig.~\ref{fig:nanodiamonds}(b). The zero-phonon line (ZPL) may be less well-defined for trapped nanodiamonds than for those on a solid surface, possibly due to heating of the particle in the high vacuum environment~\cite{lai2013quenching,li2016observation}, even for the relatively weak pump beam used.

\section*{Detecting and Cooling the Particle Motion}

\begin{figure}[ht]
\centerline{\includegraphics[width=1\columnwidth]{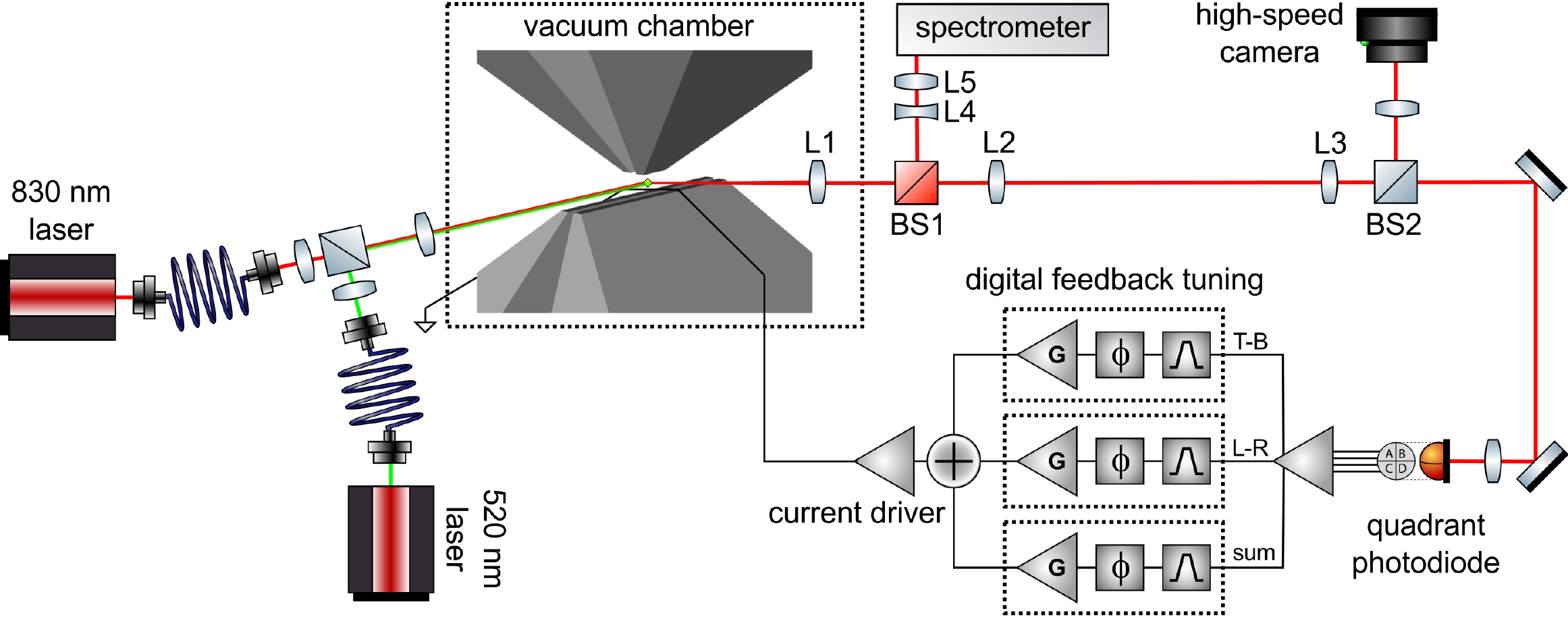}}
\caption{Optical setup of experiment for NV centre photoluminescence collection and feedback cooling. The particle can be illuminated with an $830$~nm laser for position detection and $520$~nm laser for driving NV centre photoluminescence, each coupled through a single-mode fibre. Scattered light and photoluminescence are collected by aspheric lens L1 (LightPath 355390-B), after which the scattered $830$~nm light and the photoluminescence are separated by dichroic beamsplitter BS1. The photoluminescence passes through lenses L4 and L5 to compensate for the chromatic aberration of L1 and to couple the light into a multimode fibre for detection in the spectrometer, which consists of a scanning monochromator and single photon counting module. The scattered $830$~nm light is relayed by L2 and L3 before being imaged onto the high-speed camera (Mikrotron MC1364) for recording the motion and a quadrant photodiode (Hamamatsu S5980) for feedback cooling with split vertical ($T-B=\mathrm{top}-\mathrm{bottom}$), axial ($L-R=\mathrm{left}-\mathrm{right}$) and transverse (sum) signals.}
\label{fig:optical_setup}
\end{figure}	

At rough vacuum (over $10^{-3}$~mbar), we expect the motion of a particle in the magneto-gravitational trap to be highly damped, and the temperature of that motion to be in thermal equilibrium with the gas surrounding it. Under high vacuum ($7\times 10^{-8}$~mbar), the damping time can be minutes or longer, and the temperature of the mechanical motion of the particle is largely disconnected from room temperature. 

In order to cool or measure the temperature of the particle motion, the movement of the particle must be accurately tracked. We use two detectors to observe the particle motion: a high-speed camera and a quadrant silicon PIN photodiode. In either case, the particle is illuminated from the axial direction with light from an $830$~nm diode laser (intensity $\approx$ 1~nW/$\upmu$m$^2$), and the scattered light is collected by an aspheric lens from the transverse direction (see Fig.~\ref{fig:optical_setup}). Only a small fraction ($8$\%) of the collected light is imaged onto the camera; the remainder is imaged on the quadrant detector.

Due to the low stiffness of the trap, the thermal motion of the particle at ambient temperature is easily observed on the camera (see Supplementary video V1). The centre-of-mass motion of the particle is analysed by recording high-speed ($496$~frames per second) images of the scattered light and post-processing the images (using trackpy~\cite{trackpy}, which implements a widely-used particle tracking algorithm~\cite{crocker1996methods}) to track the motion of the particle in the axial and vertical directions (for details, see Supplementary Information). The overall intensity of scattered light reaching the camera or quadrant photodiode gives some information about the transverse motion as well, but the transverse cooling is typically less effective due to the poor linearity of the overall intensity with respect to the transverse displacement. After determining the trajectory of the particle, the motion is analysed by one or more techniques to extract the temperature of the motion or the size of the particle. We primarily use the power spectral density (PSD)~\cite{norrelykke2011harmonic}, estimated from the Fourier transform of the motion in each direction, which can be written as a function of frequency $f$ as
\begin{equation}
\left\langle S_k(f) \right\rangle = S_0\frac{f_0^2\Gamma_{f0}}{\left(f_0^2 -f^2\right)^2 + f^2 \Gamma_{f0}^2}, 
\label{eq:psd_formB}
\end{equation}
where the undamped frequency $f_0$ and the damping rate $\Gamma_{f0}$ are both in units of Hz. $S_0$ is proportional to the mean square displacement of the motion, and can be written as
\begin{equation}
S_0 = \frac{k_B T}{2\pi^3 mf_0^2},
\end{equation}
where $k_B$ is the Boltzmann constant, $T$ is the temperature of the degree of freedom being measured, and $m$ is the mass of the particle.

\begin{figure}[ht]
\centerline{\includegraphics[width=1.0\columnwidth]{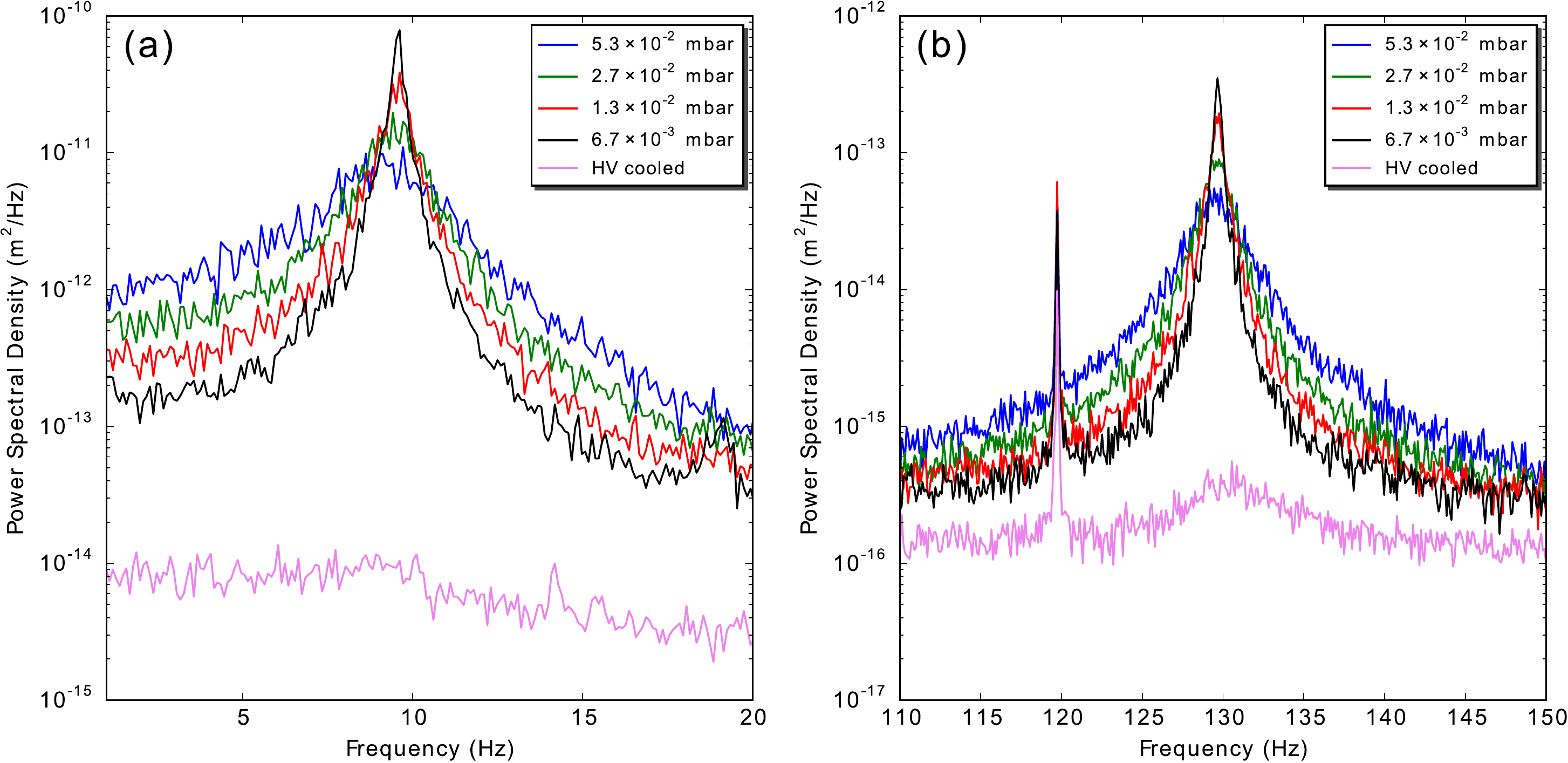}}
\caption{Results of feedback cooling, showing the power spectral density of uncooled and cooled (a) axial and (b) vertical centre-of-mass motions. At rough vacuum, the motions are in thermal equilibrium with the ambient temperature gas. At high vacuum (HV) with feedback cooling, both motions can be cooled to well below ambient temperature. The peak at $120$~Hz is due to power line noise.
}
\label{fig:damping_and_cooling}
\end{figure}

Plots of the PSD of the motion of a diamond nanocrystal cluster are shown in Fig.~\ref{fig:damping_and_cooling} at several pressures. Pressures of $7\times 10^{-3}$~mbar and higher were obtained by first pumping to high vacuum and then backfilling the chamber with nitrogen. Assuming the motion is thermalised at ambient temperature ($295$~K) at pressures of $6.7\times 10^{-3}$~mbar and higher, we can fit the PSD and extract $S_0$ to measure the mass of the particle, which for the data shown gives $m=28$~pg  (averaging the mass across axial and vertical motion at all rough vacuum pressures). If the nanodiamond cluster were a densely packed sphere, this would correspond to a radius of $1.2$~$\upmu$m. Since it is likely a loosely packed cluster with significant porosity, the usual calculation of the particle size from the damping rate with a background gas~\cite{epstein1924resistance,chang2010cavity,beresnev1990motion, li2011millikelvin,gieseler2012subkelvin,kiesel2013cavity,rahman2016burning} would not be meaningful.

The motion of the particle can be cooled with feedback of the detected motion using cold damping~\cite{milatz1953reduction}, which, in an ideal system, can be used to cool to the ground quantum state~\cite{courty2001quantum,rodenburg2016quantum}. In our experiment, the position of the particle in the $y$ and $z$ directions is tracked on the quadrant photodiode. The signals are then digitally filtered through bandpass filters, amplified, and phase shifted to minimise noise and to make them effectively proportional to the velocity of the particle. Finally, these signals are used to drive a current (typically $< 100$~mA) through a small wire which runs across the trap pole pieces (see Fig.~\ref{fig:trap_illustration}(b)) to apply a damping force to the particle through the diamagnetism of the trapped particle. Although all the feedback drive goes through a single wire in the trap, the motions of the particle in the trap are well-separated in frequency, so each degree of freedom is only significantly affected by the feedback of its own motion. We typically apply some cooling in the transverse direction based on the total intensity of light reaching the quadrant photodiode, but we do not analyse it due to poor linearity and lack of a meaningful calibration of the motion.

As the gas pressure is decreased, the damping rate of the motion of a trapped particle decreases, as can be seen by the increase in width of the resonance peaks in Fig.~\ref{fig:damping_and_cooling}. At high vacuum, the damping is greatly reduced, and the centre-of-mass motion is only weakly coupled to the environment. When manually optimised feedback cooling is added, the damping rate increases and the peak height decreases, as expected. Without analysis, the cooled particle appears to be stationary in high-speed videos (see Supplementary video V2). Assuming the mass of the particle is the same as in the damped ambient-temperature measurements, PSD analysis shows the axial motion of the particle is cooled to $600$~mK before the resonance peak becomes undetectable, while the vertical motion is cooled to $3$~K with the strongest feedback. The lowest recordable temperature for the axial motion is limited by the noise in the high-speed images used for analysis, while the vertical cooling is limited by need to avoid broadening the resonance too far, as power line noise at $120$~Hz would interfere with cooling. See Supplementary Table S1 for the full results.

The primary reason for using a diamond nanocrystal cluster instead of a single diamond nanocrystal is that the clusters remain trapped even at all relevant temperatures, allowing simple calibration of the particle size and cooling. A smaller particle can be knocked out of the trap by the thermal energy at ambient temperature, and would require cold damping to keep it trapped. Such weakly trapped particles can still be loaded at room temperature by taking advantage of the low vapor pressure of DBS. This can begin by loading a large DBS droplet containing one nanodiamond, which is easily trapped at ambient temperature. The chamber can then be evacuated and the centre-of-mass motion of the droplet cooled before the DBS has evaporated. After complete evaporation of the DBS, the diamond nanocrystal would be left in the trap under high vacuum.

\section*{Conclusions}

We have demonstrated that clusters of diamond nanocrystals can be trapped in high vacuum in a magneto-gravitational trap. The identity of the clusters was confirmed by their photoluminescence spectrum, which was taken under high vacuum while cooling the axial and vertical mechanical motion. The axial motion could be cooled from ambient temperature to $600$~mK by optical detection and magnetic feedback, while the vertical motion could be cooled to $3$~K by the same technique. The unique features of this trap include large trap size, weak restoring force, and the use of only static magnetic fields and gravity to construct the trap. The large trap size and weak restoring force make cooling relatively simple. The use of only static fields eliminates the heating problems in optical traps, making trapping in high vacuum straightforward. While heating from the pump laser used to drive the NV centres and the illumination laser used to track the particle motion cause some heating, the power levels are many orders of magnitude less than is typically required for optical trapping. With continued improvements in cooling and a reduction in the particle size, the system may be used for many of the experiments in quantum mechanics originally envisioned for optically trapped nanodiamonds. 

\bibliography{mag_trap_intro_paper_bbl}

\section*{Acknowledgements}
We thank M. V. Gurudev Dutt for invaluable discussions and for use of laboratory resources. This material is based upon work supported by the National Science Foundation under Grant No. 1540879. P.J. was supported in part by the DOE Office of Basic Energy Sciences (DE-SC 0006638).

\section*{Author contributions statement}
B.D. conceived the experiments, J.-F.H., P.J., and C.W.L. conducted the experiments, and J.-F.H. analysed the results. All authors wrote and reviewed the manuscript.

\section*{Additional information}
The authors declare no competing financial interests.


\newpage
{\center \appendix{\Large \bf{Supplementary Information}}}

\addto\captionsenglish{%
	\renewcommand{\tablename}{Supplementary Table}
	\renewcommand{\figurename}{Supplementary Figure}
}

\section*{Multipole Expansion of the Trap Magnetic Scalar Potential}
Since there are no currents in the trapping region, the magnetic field can be derived from a magnetic scalar potential $\Phi_M$ which must satisfy Laplace's equation,
\begin{align}
	\nabla^2\Phi_M = 0.
\end{align}
The general solution for $\Phi_M$ can be expanded in spherical harmonics as
\begin{align}
	\Phi_M(r,\theta,\phi) = \sum_{l=0}^{\infty}\sum_{m=-l}^l \left[ A_{lm}r^l + B_{lm}r^{-(l+1)} \right]Y_l^m(\theta, \phi),
\end{align}
where $r$ is the radial distance, $\theta$ is the polar angle, and $\phi$ is the azimuthal, $Y_l^m(\theta, \phi)$ is a complex spherical harmonic, and $A_{lm}$ and $B_{lm}$ are expansion coefficients. For a finite field throughout the trapping region, we must have $B_{lm}=0$. Based on the trap symmetry, the potential profile of the trap, and the axes designations ($x$ for the transverse direction, $y$ for the vertical direction, and $z$ for the axial direction), three real spherical harmonics are chosen to approximate the potential. Expressed in Cartesian coordinates, they are:
\begin{align}
\label{eq:RY2-2} Y_{2,-2}\left(x, y, z\right) &= i\sqrt{\frac{1}{2}}\left( Y_{2}^{-2}\left(x, y, z\right) - Y_{2}^{2}\left(x, y, z\right) \right) =  \frac{1}{2} \sqrt{\frac{15}{\pi}} \frac{xy}{r^2} \\
\label{eq:RY31} Y_{3,1}\left(x, y, z \right) &= \sqrt{\frac{1}{2}}\left( Y_{3}^{-1}\left(x, y, z\right) - Y_{3}^{1}\left(x, y, z\right) \right) =  \frac{1}{4} \sqrt{\frac{21}{2\pi}} \frac{x(4z^2-x^2-y^2)}{r^3} \\
\label{eq:RY4-4} Y_{4,-4}\left(x, y, z \right) &= i\sqrt{\frac{1}{2}}\left( Y_{4}^{-4}\left(x, y, z\right) - Y_{4}^{4}\left(x, y, z\right) \right) =  \frac{3}{4} \sqrt{\frac{35}{\pi}} \frac{xy(x^2-y^2)}{r^4}.
\end{align}
Each real spherical harmonic contributes a corresponding term to the magnetic scalar potential:
\begin{align}
	\Phi_{2,-2}\left(x, y, z\right) &= \left(\frac{a_2 y_0}{2}\right)\left(\frac{1}{y_0}\right)^2r^2 Y_{2,-2}(x,y,z) \\
\Phi_{31}\left(x, y, z\right) &= \left(\frac{a_3 y_0}{3}\right)\left(\frac{1}{y_0}\right)^3r^3 Y_{3,1}(x,y,z) \\
\Phi_{4,-4}\left(x, y, z\right) &= \left(\frac{a_4 y_0}{4}\right)\left(\frac{1}{y_0}\right)^4r^4 Y_{4,-4}(x,y,z),
\end{align}
where $a_2, a_4$, and $a_3$ are the coefficients to be solved using the observed centre-of-mass oscillation frequencies, and $y_0$ is a length scale, chosen to be the vertical distance from the geometrical centre of the trap to the upper or lower pole pieces. The arrangement of the factors in front of each term is such that $a_2, a_4$, and $a_3$ will be in units of Tesla. The magnetic scalar potential $\Phi_M$ can then be used to calculate the magnetic field as in the main text.

\section*{Trajectory Tracking}
A microscope slide with tick marks at 10~$\upmu$m intervals is imaged onto the high-speed camera through the same optical path as used for imaging trapped particles. This results in a calibration of 0.259~$\upmu$m per pixel.

The high speed camera images are analysed using trackpy, which determines the position of the particle in each frame. In approximately 1\% of the frames, the detected light is too dim for trackpy to locate the particle. In these cases, we replace the location of the particle with the average location of the particle in all frames where the particle could be located. Increasing the threshold brightness for finding the particle such that the number of frames rejected increases by a factor of two or more has only a small impact on the reported results; in particular the measured cooled temperature of the centre-of-mass motion appears to be biased up by the loss of data.

\section*{Trapped Particle in Motion}
Supplementary video V1 is an animated .gif file of the high-speed images at 0.053~mbar. Supplementary video V2 shows the cooled motion at high vacuum. The frame rate in the videos has been slowed by a factor of ten to improve visibility of the motion.

\section*{Centre-of-Mass Motion Data}
Supplementary Table~\ref{table:data} summarises the full centre-of-mass motion results from one diamond nanocrystal cluster at several pressures. Statistical uncertainties are given for each measurement, which are calculated from the standard deviation of each PSD point over 30 trials. The mass of the particle $m$ and the oscillation frequencies ($\omega_x$, $\omega_y$, and $\omega_z$) reported in the main text are weighted averages of the values calculated at rough vacuum and ambient temperature, but the final uncertainties are dominated by the systematic error which can be seen in the variation of those values with background gas pressure. The cooled temperature uncertainties are also larger than the purely statistical errors in the table due to systematic uncertainty in $m$.

\begin{table}[htb]
  \centerline{
\begin{tabular}{lcccccc}
\hline\hline
Pressure (mbar) & Direction &  $f_0$ (Hz)         & $\Gamma_{f0}$ (Hz) & $S_0$ ($\upmu$m$^2$) & $m$ (pg) & $T$ (K)    \\
\hline
$5.3\times 10^{-2}$ & Axial           &	$9.64\pm0.02$  	   & $3.39\pm0.05$     & $24.5\pm0.3$    	  & $28.8\pm0.4$  			& assumed 295\\
$5.3\times 10^{-2}$ & Vertical        &	$129.53\pm0.02$    & $3.56\pm0.06$     & $0.151\pm0.002$  	  & $25.9\pm0.3$  			& assumed 295\\
$5.3\times 10^{-2}$ & Transverse      &	$104.03\pm0.04$    & $3.2\pm0.1$       & n/a       			  & n/a           			& assumed 295\\
\hline
$2.7\times 10^{-2}$ & Axial           &	$9.56\pm0.01$  	   & $1.67\pm0.03$     & $23.3\pm0.3$     	  & $30.8\pm0.4$ 			 & assumed 295\\
$2.7\times 10^{-2}$ & Vertical        &	$129.56\pm0.02$    & $1.75\pm0.04$     & $0.146\pm0.002$  	  & $26.9\pm0.4$  			& assumed 295\\
$2.7\times 10^{-2}$ & Transverse      &	$104.03\pm0.02$    & $1.66\pm0.05$     & n/a         		  & n/a           			& assumed 295\\
\hline
$1.3\times 10^{-2}$ & Axial           &	$9.58\pm0.01$      & $0.87\pm0.02$     & $23.5\pm0.5$         & $30.5\pm0.6$  			& assumed 295\\
$1.3\times 10^{-2}$ & Vertical        &	$129.65\pm0.01$    & $0.91\pm0.03$     & $0.149\pm0.004$      & $26.3\pm0.7$  			& assumed 295\\
$1.3\times 10^{-2}$ & Transverse      &	$104.05\pm0.01$    & $0.87\pm0.02$     & n/a        	      & n/a           			& assumed 295\\
\hline
$6.7\times 10^{-3}$ & Axial            &	$9.57\pm0.01$      & $0.40\pm0.01$     & $25.9\pm0.7$         & $27.7\pm0.7$  			& assumed 295\\
$6.7\times 10^{-3}$ & Vertical         &	$129.66\pm0.01$    & $0.47\pm0.02$ 	   & $0.150\pm0.006$      & $26\pm1$      			& assumed 295\\
$6.7\times 10^{-3}$ & Transverse       &	$104.10\pm0.01$    & $0.44\pm0.02$ 	   & n/a          		  & n/a          		    & assumed 295\\
\hline
HV cooled & Axial 		  &	$10.3\pm0.2$       & $10.6\pm0.6$      & $0.046\pm0.003$ 	  & assumed $27.8$ 			& $0.60\pm0.05$  \\
HV cooled & Vertical 	  & $130.72\pm0.08$    & $6.3\pm0.3$       & $(1.5\pm0.1)\times10^{-3}$  & assumed $27.8$   & $3.2\pm0.2$    \\
HV cooled & Transverse & $105.00\pm0.01$ & n/a    & n/a & n/a                 & n/a     \\
\hline\hline
\end{tabular}}
\renewcommand\thetable{S1}
\caption{\label{table:data}Details of centre-of-mass motion for a single diamond nanocrystal cluster. High vacuum (HV) indicates a pressure of $7\times 10^{-8}$~mbar. All reported uncertainties are statistical.}
\end{table}

\end{document}